\newcommand{\printTitle}{A Note on the Output of a Coordinate-Exchange Algorithm for Optimal Experimental Design}
\newcommand{\printWho}{\title{\bf \printTitle}
\date{}
\author{Arno Strouwen*\\
	{\small arno.strouwen@kuleuven.be}\\
	{\small ORCID: 0000-0001-8607-4091}\\
	{\small Faculty of Bioscience Engineering}\\{\small Department of Biosystems}\\{\small KU Leuven}\vspace{-0.3cm}\\
	$\mbox{ }$\\
	Peter Goos\\
	{\small peter.goos@kuleuven.be}\\
	{\small ORCID: 0000-0002-3854-6506}\\
	{\small Faculty of Bioscience Engineering}\\{\small Department of Biosystems}\\{\small KU Leuven}\vspace{-0.3cm}\\
	\\
	{\small Faculty of Business and Economics }\\{\small Department of Engineering Management}\\{\small University of Antwerp}}
\maketitle
\maketitle
\maketitle
}

\newcommand{\printAbstract}{The coordinate-exchange algorithm is commonly used to construct optimal experimental designs. Every execution of the coordinate-exchange algorithm produces a new, seemingly random, order of the selected design points. In this short communication, we study the order of the design points produced by the algorithm and conclude that certain orders appear much more often than others. As a result, an explicit randomization step of the design points is required before conducting an experiment using a design produced by a coordinate-exchange algorithm.}
\newcommand{\printKeywords}{coordinate-exchange algorithm, D-optimality, optimal experimental design, randomization, run order.}

\documentclass[12pt]{article}

\usepackage[T1]{fontenc}
\usepackage[latin1]{inputenc}

\usepackage{fbb}
\usepackage{montserrat}
\usepackage[adobe-utopia]{mathdesign}

\def\spacingset#1{\renewcommand{\baselinestretch}%
	{#1}\small\normalsize} \spacingset{1}
\usepackage{natbib}
\usepackage[monochrome]{color}
\usepackage{amsmath}

\usepackage{esdiff}

\usepackage[ruled,lined]{algorithm2e}
\usepackage{hyperref}
\usepackage{graphicx,psfrag,epsf}
\usepackage{float}
\usepackage{subcaption}

\begin{document}
\printWho

\vspace{0.5cm}
\noindent\fbox{\parbox{\textwidth}{%
\small Published in \textit{Chemometrics and Intelligent Laboratory Systems}, 192, 103819, 2019.\\
DOI: \url{https://doi.org/10.1016/j.chemolab.2019.103819}\\[0.3em]
\textcopyright~2019. This manuscript version is made available under the CC-BY-NC-ND 4.0 license\\
\url{https://creativecommons.org/licenses/by-nc-nd/4.0/}
}}
\vspace{0.5cm}

\begin{abstract}
	\noindent \printAbstract \\
\end{abstract}

\noindent%

\newpage
\spacingset{1.5}

In recent years, optimal experimental design has gained popularity due to its flexibility in terms of the number of observations, experimental region and blocking \citep{goos}. For instance, \cite{akkermans} use optimal experimental design to plan excipient compatibility studies in the pharmaceutical industry, \cite{jeirani} apply optimal experimental design to determine the optimum aqueous phase formulation of a microemulsion, and \cite{mancenido} study optimal design of mixture experiments for logistic regression models.\\

One of the most commonly used algorithms to generate optimal experimental designs is the coordinate-exchange algorithm of \cite{meyer}, which has been implemented in statistical software packages such as JMP and Design Expert. The original algorithm was intended for constructing optimal designs for completely randomized experiment{\color{red}s}, in which the experimental observations are independent. In recent years, the algorithm has been adapted to cope with split-plot, split-split-plot and strip-plot design \citep{jones2,jones,arnouts,arnouts2,trinca}, deal with Bayesian optimality criteria for experimental design \citep{kessels,mylona,mylona2}, and produce foldover designs \citep{errore}. The algorithm has also been studied in detail by \citet{cuervo}.\\

In this short communication, we study the order of the design points produced by the coor\-di\-nate-exchange algorithm for completely randomized experiments. Our study is inspired by the fact that the order seems random, as a result of which experimenters may not randomize the order of the test combinations suggested by the coordinate-exchange algorithm. The pseudo-code of the coordinate-exchange algorithm for completely randomized experiments is given in Figure \ref{pseudo1}. In the pseudo-code, the symbols $n$ and $m$ represent the number of observations and the number of factors, respectively.\\
\begin{figure}[H]
\begin{algorithm}[H]
\label{pseudo1}
design = random feasible design\;
improvement = yes\;
\While{improvement == yes}{
	improvement = no\;
	\For{$i = 1$ to $n$}{
	  \For{ $j = 1$ to $m$}{
		  optimize $(i,j)$th coordinate of the design {\color{red}using an interior point algorithm}\;
		  \If{$(i,j)$th coordinate was changed}{
			improvement = yes\;
		  }
	  }
	}
}
	\end{algorithm}
	\caption{{\color{red}Single execution of the} coordinate-exchange algorithm}
\end{figure}

The coordinate-exchange algorithm starts by randomly selecting a level for each factor from the continuous uniform distribution on the $[-1,1]$ interval, for each observation. {\color{red}In our implementation, we generated random numbers using the Mersenne twister described in \cite{matsumoto}, as implemented by Matlab.} The resulting initial experimental design is then improved factor level by factor level, starting with the level of the first factor at the first observation and ending with the level of the last factor at the last observation. Each factor level is optimized using a one-dimensional continuous optimizer, keeping all other factor levels constant. For the results we report in this short communication, we utilized an interior point algorithm as implemented in the Matlab function fmincon, and described in \citet{byrd,byrd2} and \citet{waltz}. {\color{red} The interior point method is a popular choice in recent implementations of the coordinate exchange algorithm \citep{ruseckaite,huang}. The orignal coordinate-exchange algorithm of \cite{meyer}, however, used grid optimization rather than a continuous optimizer.} The coordinate-exchange algorithm continues until a whole pass through the entire design does not yield a single improvement {\color{red}anymore}.\\

To make our points, we focus on a relatively simple example. More specifically, we study the generation of a D-optimal design for a specific linear regression model involving two quantitative factors, $x_1$ and $x_2$. The model involves the main effects of both factors, their interaction effect and the quadratic effect of $x_1$. Hence, the model is given by
\begin{equation}\label{modelleke}
Y_i = \beta_0 + \beta_1x_{i1} + \beta_2x_{i2} + \beta_{11}x_{i1}^2 + \beta_{12}x_{i1}x_{i2} + \varepsilon_i,
\end{equation}
where $Y_i$ represents the response at the $i$th observation, $x_{i1}$ and $x_{i2}$ represent the levels of the factors $x_1$ and $x_2$ at that observation, and $\varepsilon_i$ is the random error term. The model involves five unknown parameters, $\beta_0$, $\beta_1$, $\beta_2$, $\beta_{11}$ and $\beta_{12}$. D-optimal designs maximize the determinant of the information matrix about these five parameters \citep{goos}. The only D-optimal design for model \eqref{modelleke} involves the four factor level combinations of the $2^2$ factorial design as well as the combinations $(0,-1)$ and $(0,1)$. The order in which the factor level combinations are tested does not impact the optimality of the experimental design, but it is recommended in design of experiments textbooks to carry out all tests in a random order. This is because randomization is one of the main experimental design principles \citep{mont}, and it is crucial to justify the traditional statistical analyses of the experimental data \citep{coxdr}.\\

When looking at the output of a coordinate-exchange algorithm, it is tempting to believe that the algorithm outputs the optimal design points in a random order. Table \ref{ordersof3starts} shows the order of the six optimal design points produced by three runs of the coordinate-exchange algorithm. In each case, the order of the six D-optimal factor level combinations is different. This suggests that the ordering of the factor level combinations could be considered random.\\

\begin{table}
\caption{Run orders produced by three executions of the coordinate-exchange algorithm for model~\eqref{modelleke}.}
\begin{center}
\label{ordersof3starts}
\begin{tabular}{|c|rr|rr|rr|}
\hline 
Obs. & \multicolumn{2}{|c|}{Run 1} & \multicolumn{2}{|c|}{Run 2} & \multicolumn{2}{|c|}{Run 3}\\
& $x_1$ & $x_2$ & $x_1$ & $x_2$& $x_1$ & $x_2$\\
\hline
$1	$&$	0	$&$	1	$&$	1	$&$	1	$&$	0	$&$	-1	 $\\
$2	$&$	-1	$&$	1	$&$	-1	$&$	-1	$&$	1	$&$	1	 $\\
$3	$&$	1	$&$	-1	$&$	-1	$&$	1	$&$	1	$&$	-1	 $\\
$4	$&$	0	$&$	-1	$&$	1	$&$	-1	$&$	-1	$&$	-1	 $\\
$5	$&$	1	$&$	1	$&$	0	$&$	1	$&$	-1	$&$	1	 $\\
$6	$&$	-1	$&$	-1	$&$	0	$&$	-1	$&$	0	$&$	1	 $\\
\hline
\end{tabular}
\end{center}
\end{table}

To test whether the design points' ordering is indeed random, we conducted a computational experiment involving a large number of executions of the coordinate-exchange algorithm, and we recorded the order in which the six optimal design points occur in the outputs of these executions. As there are six distinct factor level combinations in the D-optimal design, there are $6!=720$ possible orders. In the event the coordinate-exchange algorithm's output would be random, these 720 orders should appear roughly equally often. After executing the coordinate-exchange algorithm $56,000$ times, we observed that each of the 720 orders had been produced at least once and we stopped our computational experiment.\\

\begin{figure}
	\centering
	\includegraphics[width=0.8\textwidth]{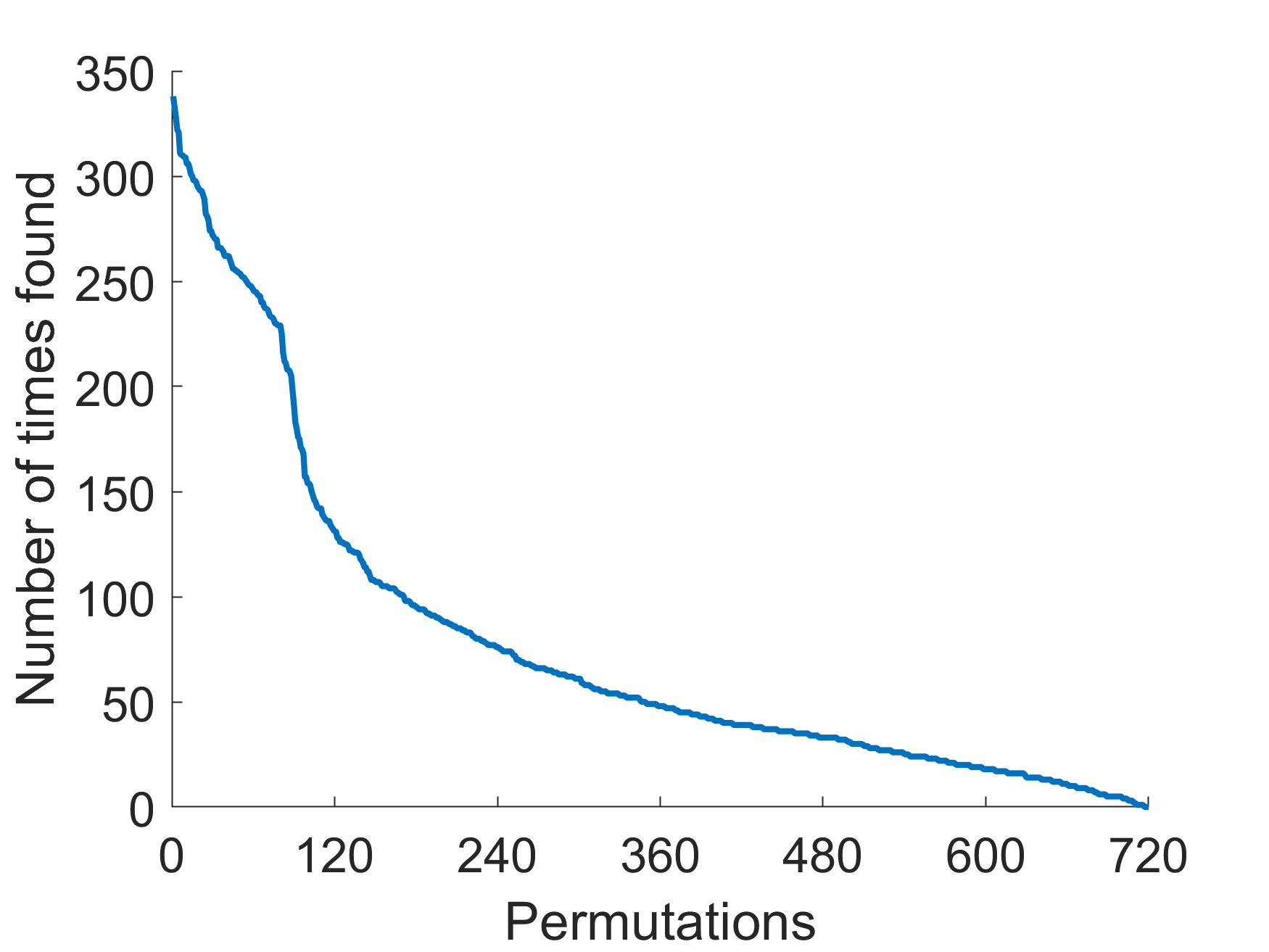}
	\caption{Frequency distribution of the 720 different orders of the six D-optimal factor level combinations.}
	\label{figPermutations}
\end{figure}

When studying the results, it turned out that the 720 orders did not at all appear equally often in the outputs of the $56,000$ executions of the algorithm. Figure \ref{figPermutations} shows the frequencies with which each of the 720 orders of the six D-optimal factor level combinations were found. The figure shows that certain orders were found more than 250 times, while others were encountered rarely. The figure therefore suggests that the orders produced by the coordinate-exchange algorithm are not at all random. This is confirmed by Pearson's chi-squared test, which produces a $p$ value much smaller than $0.001$. We can therefore safely reject the hypothesis that the orders are random.\\

We also studied the orders that occurred most frequently. Generally, the test combinations of the $2^2$ factorial design appear first, while the two factor level combinations involving the middle level of the first factor appear in the second half of the design. Table \ref{bestworst} shows the six orders that occurred most frequently in the 56,000 executions of the coordinate-exchange algorithm. All six orders start with three of the four factorial points. For two of these orders, the fourth design point is the fourth factorial point. For the four other orders, the fourth design point involves the middle level of factor $x_1$. Overall, 12,779 of the 56,000 executions of the algorithm produced a run order starting with the four factorial points, while 7608 of the 56,000 executions produced a run order in which the four factorial points appear in positions 1, 2, 3 and 5. As there are $48$ run orders starting with the four factorial points as well as $48$ run orders starting with three of the factorial points, there is a minority of $96$ run orders {\color{red}which are} produced much more frequently than the $624$ other run orders.
The exact frequencies are $36\%$ for the 96 run orders similar to those in Table \ref{bestworst} and $64\%$ for the remaining $624$ others.\\

\begin{table}
	\caption{Most common run orders produced by the coordinate-exchange algorithm for model~\eqref{modelleke}.}
	\begin{center}
		\label{bestworst}
		\begin{tabular}{|c|rr|rr|rr|rr|rr|rr|}
			\hline 
			Obs. & \multicolumn{2}{|c|}{1} & \multicolumn{2}{|c|}{2} & \multicolumn{2}{|c|}{3}&
			\multicolumn{2}{|c|}{4} & \multicolumn{2}{|c|}{5}&
			\multicolumn{2}{|c|}{6}
			\\
			& $x_1$ & $x_2$ & $x_1$ & $x_2$& $x_1$ & $x_2$ & $x_1$ & $x_2$ & $x_1$ & $x_2$& $x_1$ & $x_2$\\
			\hline
			$1	$&$	-1	$&$	1	$&$	1	$&$	1	$&$	1	$&$	1
			    $&$	-1	$&$	1	$&$	-1	$&$	-1	$&$	1	$&$	1 $\\
			$2	$&$	1	$&$	1	$&$	-1	$&$	1	$&$	-1	$&$	1	 	
			    $&$	1	$&$	1	$&$	1	$&$	-1	$&$	1	$&$	-1 $\\
			$3	$&$	1	$&$	-1	$&$	-1	$&$	-1	$&$	1	$&$	-1
			    $&$	-1	$&$	-1	$&$	1	$&$	1	$&$	1	$&$	-1 $\\	 
			$4	$&$	-1	$&$	-1	$&$	0	$&$	1	$&$	-1	$&$	-1	 
			    $&$	0	$&$	1	$&$	0	$&$	-1	$&$	0	$&$	1 $\\
			$5	$&$	0	$&$	-1	$&$	1	$&$	-1	$&$	0	$&$	1
			    $&$	1	$&$	-1	$&$	-1	$&$	1	$&$	-1	$&$	-1 $\\
			$6	$&$	0	$&$	1	$&$	0	$&$	-1	$&$	0	$&$	1
			    $&$	0	$&$	-1	$&$	0	$&$	1	$&$	0	$&$	-1 $\\
			\hline
		\end{tabular}
	\end{center}
\end{table}


We can explain the dominance of the factor level combinations of the $2^2$ factorial design in the first half of the D-optimal design as follows. The D-optimality criterion favors extreme points of the experimental region because they have a large positive impact on the D-optimality criterion. Therefore, when optimizing the levels of the factors in the first couple of observations of an experiment, the coordinate-exchange algorithm will tend to select the extreme levels for the various experimental factors. This guarantees a precise estimation of the linear main effects $\beta_1$ and $\beta_2$ in model~\eqref{modelleke}, and of the interaction effect $\beta_{12}$. Selecting the factorial points therefore leads to the largest leap in D-optimality. The middle level of the first factor is only useful for estimating the quadratic effect $\beta_{11}$. So, that factor level is important for fewer model parameters than the extreme levels $-1$ and $1$. It is also known that D-optimal designs use the middle level less frequently than the extreme factor levels for models involving quadratic effects \citep{atdoto}.\\
\\
\\
{\color{red}To make sure that our result does not depend on the specific random number generator we used to create the initial design in the coordinate-exchange algorithm, we tested two different initial seeds of the Mersenne twister of \cite{matsumoto} and we tried a different random numer generator (more specifically the multiplicative congruential generator described in \cite{park} and implemented by Matlab as well). All these tests also led to a $p$ value smaller than $0.001$ for Pearson's chi-squared test, and thus to a rejection of the hypothesis that the run order produced by the algorithm is random. We reached similar conclusions when considering A-optimality instead of D-optimality as a criterion to select the design. The non-randomness of the run order produced by the coordinate-exchange algorithms thus is neither caused by the random number generation nor by the optimality criterion considered. Instead, it is caused by the sequential, row-by-row, way in which the levels of the factors are optimized.
}
\\
\\
The practical implication of the result that the order of the factor level combinations produced by a coordinate-exchange algorithm are not random, is that a proper randomization of the full set of combinations is required before carrying out the experiment. In the event a coordinate-exchange algorithm is used for designing blocked or split-plot experiments, it will be required to randomize the blocks/whole plots as well as the observations within blocks/whole plots after the design has been generated. For other multi-stratum designs, a similar randomization will be necessary.\\

{\color{red}One approach to ensure that the output of a coordinate-exchange algorithm has a random run order is to adapt the algorithm so that it visits the rows of the experimental design in a random order, rather than sequentially from row 1 to row $n$. However, this is more cumbersome than randomizing the original algorithm's output}.
 
\section*{Acknowledgements}
Author Arno Strouwen is a PhD fellow Strategic Basic Research (SB) of the Fund for
Scientific Research, Flanders (FWO), project 1S58717N.
\small
\bibliography{refs}

@article{akkermans,
title = "Optimal design of experiments for excipient compatibility studies",
journal = "Chemometrics and Intelligent Laboratory Systems",
volume = "171",
pages = "125 - 139",
year = "2017",
author = "Wannes G.M. Akkermans and Hans Coppenolle and Peter Goos",
}

@book{atdoto,
author = "Atkinson, A. C. and Donev, A. N. and Tobias, R. D.",
title = "Optimum Experimental Designs, with SAS",
publisher = "Oxford: Oxford University Press",
year = "2007"
}

@article{coxdr,
author = {Cox, D. R.},
title = {Randomization in the Design of Experiments},
journal = {International Statistical Review},
volume = {77},
pages = {415--429},
year = {2009}
}

@article{huang,
author = {Huang, Yuanzhi and Gilmour, Steven G. and Mylona, Kalliopi and Goos, Peter},
title = {Optimal design of experiments for non-linear response surface models},
journal = {Journal of the Royal Statistical Society: Series C (Applied Statistics)},
volume = {68},
pages = {623--640},
year = {2019}
}

@article{kessels,
author = {Roselinde Kessels and Bradley Jones and Peter Goos and Martina Vandebroek},
title = {An Efficient Algorithm for Constructing {B}ayesian Optimal Choice Designs},
journal = {Journal of Business \& Economic Statistics},
volume = {27},
pages = {279-291},
year  = {2009}
}

@book{mont,
author = "Montgomery, D. C.",
title = "Design and Analysis of Experiments",
publisher = "New York: Wiley",
year = "2012"
}

@article{mylona,
author = {Kalliopi Mylona and Peter Goos and Bradley Jones},
title = {Optimal Design of Blocked and Split-Plot Experiments for Fixed Effects and Variance Component Estimation},
journal = {Technometrics},
volume = {56},
pages = {132-144},
year  = {2014},
}

@article{mylona2,
author = {Kalliopi Mylona and Steven G. Gilmour and Peter Goos},
title = {Optimal Blocked and Split-Plot Designs Ensuring Precise Pure-Error Estimation of the Variance Components},
journal = {Technometrics},
volume = {61},
pages = {To appear},
year  = {2019}}

@article{jeirani,
title = "The optimal mixture design of experiments: Alternative method in optimizing the aqueous phase composition of a microemulsion",
journal = "Chemometrics and Intelligent Laboratory Systems",
volume = "112",
pages = "1--7",
year = "2012",
author = "Zahra Jeirani and Badrul Mohamed Jan and Brahim Si Ali and Ishenny Mohd. Noor and See Chun Hwa and Wasan Saphanuchart",
}

@article{mancenido,
title = "Comparing {D}-optimal designs with common mixture experimental designs for logistic regression",
journal = "Chemometrics and Intelligent Laboratory Systems",
volume = "187",
pages = "11--18",
year = "2019",
author = "Michelle V. Mancenido and Rong Pan and Douglas C. Montgomery and Christine M. Anderson-Cook",
}

@article{arnouts,
  title={Design and analysis of industrial strip-plot experiments},
  author={Arnouts, Heidi and Goos, Peter and Jones, Bradley},
  journal={Quality and Reliability Engineering International},
  volume={26},
  pages={127--136},
  year={2010},
  publisher={Wiley Online Library}
}

@article{arnouts2,
  title={Staggered-level designs for response surface modeling},
  author={Arnouts, Heidi and Goos, Peter},
  journal={Journal of Quality Technology},
  volume={47},
  pages={156--175},
  year={2015},
  publisher={Taylor \& Francis}
}

@article{byrd,
  title={An interior point algorithm for large-scale nonlinear programming},
  author={Byrd, Richard H and Hribar, Mary E and Nocedal, Jorge},
  journal={SIAM Journal on Optimization},
  volume={9},
  pages={877--900},
  year={1999},
  publisher={SIAM}
}

@article{byrd2,
  title={A trust region method based on interior point techniques for nonlinear programming},
  author={Byrd, Richard H and Gilbert, Jean Charles and Nocedal, Jorge},
  journal={Mathematical Programming},
  volume={89},
  pages={149--185},
  year={2000},
  publisher={Springer}
}

@article{cuervo,
  title={Optimal design of large-scale screening experiments: a critical look at the coordinate-exchange algorithm},
  author={Cuervo, Daniel Palhazi and Goos, Peter and S{\"o}rensen, Kenneth},
  journal={Statistics and Computing},
  volume={26},
  pages={15--28},
  year={2016},
  publisher={Springer}
}

@article{errore,
  title={Benefits and fast construction of efficient two-level foldover designs},
  author={Errore, Anna and Jones, Bradley and Li, William and Nachtsheim, Christopher J},
  journal={Technometrics},
  volume={59},
  pages={48--57},
  year={2017},
  publisher={Taylor \& Francis}
}

@book{goos,
  title={Optimal design of experiments: a case study approach},
  author={Goos, Peter and Jones, Bradley},
  year={2011},
  publisher={John Wiley \& Sons}
}

@article{jones,
  title={D-optimal design of split-split-plot experiments},
  author={Jones, Bradley and Goos, Peter},
  journal={Biometrika},
  volume={96},
  pages={67--82},
  year={2009},
  publisher={Oxford University Press}
}

@article{jones2,
  title={A candidate-set-free algorithm for generating {D}-optimal split-plot designs},
  author={Jones, Bradley and Goos, Peter},
  journal={Journal of the Royal Statistical Society: Series C (Applied Statistics)},
  volume={56},
  pages={347--364},
  year={2007},
  publisher={Wiley Online Library}
}

@article{meyer,
  title={The coordinate-exchange algorithm for constructing exact optimal experimental designs},
  author={Meyer, Ruth K and Nachtsheim, Christopher J},
  journal={Technometrics},
  volume={37},
  pages={60--69},
  year={1995},
  publisher={Taylor \& Francis Group}
}

@article{trinca,
  title={Improved split-plot and multistratum designs},
  author={Trinca, Luzia A and Gilmour, Steven G},
  journal={Technometrics},
  volume={57},
  pages={145--154},
  year={2015},
  publisher={Taylor \& Francis}
}

@article{waltz,
  title={An interior algorithm for nonlinear optimization that combines line search and trust region steps},
  author={Waltz, Richard A and Morales, Jos{\'e} Luis and Nocedal, Jorge and Orban, Dominique},
  journal={Mathematical Programming},
  volume={107},
  pages={391--408},
  year={2006},
  publisher={Springer}
}

@article{ruseckaite,
  title={Bayesian {D}-optimal choice designs for mixtures},
  author={Ruseckaite, Aiste and Goos, Peter and Fok, Dennis},
  journal={Journal of the Royal Statistical Society: Series C (Applied Statistics)},
  volume={66},
  number={2},
  pages={363--386},
  year={2017},
  publisher={Wiley Online Library}
}

@article{park,
  title={Random number generators: good ones are hard to find},
  author={Park, Stephen K and Miller, Keith W},
  journal={Communications of the ACM},
  volume={31},
  pages={1192--1201},
  year={1988},
  publisher={ACM}
}

@article{matsumoto,
  title={Mersenne twister: a 623-dimensionally equidistributed uniform pseudo-random number generator},
  author={Matsumoto, Makoto and Nishimura, Takuji},
  journal={ACM Transactions on Modeling and Computer Simulation (TOMACS)},
  volume={8},
  pages={3--30},
  year={1998},
  publisher={ACM}
}

\end{document}